\documentclass[pra]{revtex4}

\usepackage{epsfig}
\usepackage{amsmath,amssymb,amsfonts}

\newcommand{\beq}{\begin{equation}}
\newcommand{\eeq}{\end{equation}}
\newcommand{\nn}{\nonumber}
\newcommand{\al}{\alpha}
\newcommand{\ii}{{\rm i}}

\newcommand{\ket}[1]{\left|{#1}\right\rangle}
\newcommand{\bra}[1]{\left\langle{#1}\right|}

\newcommand{\tr}[1]{{\rm tr}\left\{#1\right\}}

\begin{document}

\title{A geometrical relation between symmetric operators and mutually unbiased operators}
\author{Amir Kalev}
\affiliation{Centre for Quantum Technologies, National University of Singapore, 3 Science Drive 2, 117543, Singapore,\\Center for Quantum Information and Control, University of New Mexico, Albuquerque, NM 87131-0001, USA\footnote{Current address}}
\begin{abstract}
In this work we study the relation between the set of symmetric operators and the set of mutually unbiased operators from finite plane geometry point of view. Here symmetric operators are generalization of symmetric informationally complete probability-operator measurements (SIC POMs), while mutually unbiased operators are the operator generalization of mutually unbiased bases (MUB). We also discuss the implication of this relation to the particular cases of rank-1 SIC POMs and MUB.
\end{abstract}

\maketitle
\section{Introduction}\label{intro}
The geometry of quantum states and their structure has attracted the attention of many researchers; as a representative example we refer to the book of Ref.~\cite{bengbook} and the references within. Two particular interesting sets of states, that share a common geometrical structure and symmetric features, are related to the well known mutually unbiased bases (MUB) and to the symmetric, informationally complete probability-operator measurements (SIC POMs). The former have an important role in the description of quantum systems. For example, they are related to the Principle of Complementarity of Bohr \cite{englertbook}, and to the wave-particle duality nature of quantum systems \cite{englert08}. Two bases are said to be unbiased if the transition probability from any state of one basis to any state of the second basis is independent of the chosen states. In fact, in a finite-dimensional Hilbert space, $d$, there are at most $d{+}1$ bases which are pair-wise unbiased, that is, $d{+}1$ MUB \cite{ivanovic81}. The construction of the complete set of MUB is known in prime-power dimensions \cite{ivanovic81,wootters89,tal02,klap04,durt05}. However, it is still not known whether the complete set of MUB exists in dimensions which are not prime-power.  A possible construction for an odd prime dimension, $d$, is given as follows: The first basis is the computational basis  $\{|n\rangle\}_{n{=}0}^{d{-}1}$,  composed of the $d$ orthonormal eigenstates of the generalized Pauli operator $\hat{Z}$, $\hat{Z}|n\rangle=\omega^n|n\rangle, |n{+}d\rangle{=}|n\rangle,\omega=e^{\ii\frac{2\pi}{d}}$. The other $d$ orthonormal bases are parametrized by $b{=}0,1,\ldots,d{-}1$. The kets that compose the $d$ remaining bases are given in terms of the computational basis by \cite{wootters89,tal02},
\begin{equation}\label{mub}
|m;b\rangle=\frac{1}{\sqrt d}\sum_{n=0}^{d-1}|n\rangle\omega^{\frac{b}{2}n(n-1)+mn},\;m=0,1,\ldots,d-1.
\end{equation}
We shall designate the computational basis by $b=d$, and depending on the context we may also denote the kets of the computational basis $|m\rangle$ by $|m;d\rangle$. Thus, the $d{+}1$ bases are labelled by $b=0,1,\ldots,d$. The collection of the $d(d{+}1)$ vectors, $\{\ket{m;b}\}$, has a special symmetry structure, which becomes evident by their defining property,
\begin{equation}\label{mub def}
|\!\bra{m;b}{m;b}\rangle|^2=\left\{
  \begin{array}{ll}
    1 & \quad \text{for}\;b=b'\; \text{and}\; m=m',\\
    0 & \quad \text{for}\;b=b'\; \text{and}\; m\neq m',\\
    \frac1{d} & \quad \text{for}\;b\neq b'.
  \end{array} \right.
\end{equation}

The latter, the states related to SIC POMs, has attracted the attention of scholars. These states are related related to what is known as quantum designs \cite{zauner11} In $d$-dimensional Hilbert space, this set is composed of $d^2$ normalized kets, $\{\ket{\mu}:\mu{=}0,\ldots,d^2{-}1\}$, such that the transition probability of one state to any other is independent of the chosen states,
\begin{equation}\label{sic}
|\!\!\bra{\mu}\mu'\rangle|^2=\left\{
  \begin{array}{ll}
    1 & \quad \text{for}\;\mu=\mu',\\
    \frac1{d+1} & \quad \text{for}\;\mu\neq\mu',
  \end{array} \right.
\end{equation}
These states are related to the notion of  SIC POM. In quantum theory the outcomes of a measurement are mathematically represented by positive operators, $M_j\leq0$, that sum-up to one,  ${\sum_j}{M_j}=1$. The measurement is therefore a POM. The probability to obtain an outcome is given by Born's rule, $\tr{M_j\rho}$, where $\rho$ is the statistical operator of the quantum system (the mathematical representation of the state of the system). A POM is informationally complete if any state of the system is determined completely by the outcomes' probabilities. If the set $\{\ket{\mu}\}$ exists in $d$-dimensional Hilbert space than the (unnormalized) set of projectors $\{\ket{\mu}\!\frac1{d}\!\bra{\mu}\}$ is a SIC POM. For brevity we shall refer to the set $\{\ket{\mu}\}$ as the set of SIC-states. It is still an open question whether SIC-states exist in any finite dimension. Some are known to exist, by construction, in certain dimensions, see for example \cite{renes04,scott10}. 

At a first glance the two symmetric sets, the set of MUB and the set of SIC-states, when both exist, are not related to each other. However quantum design theory provides a common formulation  for the both \cite{zauner11}.  MUB and SIC-states are affine quantum designs and regular quantum designs, respectively, of particular kinds. A geometric approach has also been studied to connect the two sets \cite{bengtsson07,appleby09}. In this approach  the projectors onto MUB and the SIC POMs are mapped to vectors in real vector space. It was shown \cite{appleby09} that the complete set of MUB is mapped to $d{+}1$ mutually orthogonal $d{-}1$ dimensional regular simplexes, while SIC POMs are mapped to $d^2{-}1$ dimensional regular simplex. A regular simplex is a generalization of an equiangular triangle to a higher-dimensional real vector space. Based on this approach, several geometrical relationships between the regular simplexes and the orthogonal simplexes have been found for the case of Heisenberg-Weyl group covariant SIC POMs (HW SIC POMs).
A HW SIC POM, or equivalently HW SIC-states, are generated from a single ket, a fiducial state, $\ket{\psi_0}$, under the action of the HW group elements,
\begin{equation}\label{hw}
\ket{\psi_0}=X^kZ^j\ket{\psi_0},\; j,k=0,\ldots,d-1.
\end{equation}
The fiducial state is chosen such that the SIC-states satisfy the defining property of a SIC POM of Eq.~(\ref{sic}). The generators of the HW group are the generalized Pauli operators $Z$, which is defined just above Eq.~(\ref{mub}), and $X=\sum_{n=0}^{d-1}{\ket{n\oplus 1}\bra{n}}$ where $\oplus$ stands for the sum modulo $d$.

The purpose of this work is to explore further connection between the set of MUB and SIC-states. We take a geometrical approach, that is based on finite projective planes geometries. Similar approach was taken in related research in \cite{wootters87,wootters06,revzen11}. We begin in the next section, Sec.~\ref{simplex}, by generalizing the concept of MUB and SIC-states, to operators. Using geometrical properties of hermitian operators in real vector space we define two sets of (hermitian) operators which we term mutually unbiased operators (MUO) and symmetric operators (SO). These sets are equivalent to the sets of MUB and SIC-states when the operators are of rank-one, i.e., projectors onto pure states. On Sec.~\ref{geometry} we present the main result of our work. We show that in prime-power dimensions the MUO and SO are related to each other in the way as point and lines are related in dual affine finite plane geometries (DAPGs). In particular when the points of the DAPG are associated with projectors onto MUB, than the the lines of the plane correspond to hermitian operator basis. While if the lines of the plane correspond to projectors onto SIC-states, then the points correspond to MU POM (note that we use MU POM in plural form), which together are IC. The later case is analyzed further as it hints on the possible structure of SIC POMs in prime-power dimensions. Some of the results of \cite{appleby09} are re-derived here using this formulation. Then, on Sec.~\ref{hw-geometry}, we show that for in prime-power dimensions, when DAPG exists, if one associate the points with MUO, then the lines of the geometry (that is the SO) are generated by the action of the HW group elements on a fiducial line. Based on this observation a condition for rank-one SO (that is, a rank-one SIC POM) is derived. Finally we close by a summary and concluding remarks on Sec.~\ref{summary}.

\section{symmetric operators, mutually unbiased operators, and simplexes}\label{simplex}
Any hermitian operator with unit trace acting on a $d$-dimensional Hilbert space of kets, ${\cal H}^d$, could be written as $\tau{=}\frac1{d}(1{+}t)$, where $t$ is a traceless hermitian  operator. Of course, $\tau$ represents a statistical operator of a quantum system (i.e., a quantum state) if and only if $\tau{\geq}{0}$. In what follows we will not be restricted to this case. Being traceless hermitian operators acting on ${\cal H}^d$, the $t$s are elements of a $(d^2{-}1)$-dimensional real vector space, ${\mathbb R}^{d^2{-}1}$. Each traceless hermitian operator $t$ is represented in ${\mathbb R}^{d^2{-}1}$ by a vector $\vec{t}$, such that $t=\sum_i({\vec t}\;)_ib_i$, where  $\{b_i:i{=}1,\ldots,d^2{-}1\}$ is a hermitian operator basis, and $(\vec{t}\;)_i$ is the $i$th component of $\vec{t}$.

\subsection{Symmetric hermitian operators}\label{simplex sub a}
Consider a regular simplex in ${\mathbb R}^{d^2{-}1}$. A regular simplex is a generalization of an equiangular triangle to a higher-dimensional real vector space. In ${\mathbb R}^{d^2{-}1}$, it is composed from $d^2$ equilength vectors, $\{\vec{s}_i:i{=}0,\ldots,d^2{-}1\}$,  with equal pair-wise scalar product, 
\beq\label{equi vec}
\vec{s}_i\cdot\vec{s}_j=\left\{
  \begin{array}{ll}
    \al & \quad \text{for}\;i=j,\\
    \al \Bigl(-\frac1{d^2-1}\Bigr)& \quad \text{for}\;i\neq j.
  \end{array} \right.
\eeq
The set $\{\vec{s}_i\}$ is called $(d^2{-}1)$-simplex. These properties ensures that $\sum_i \vec{s}_i{=}0$.

Now, associate with each vector of the simplex, $\vec{s}_i$, a traceless hermitian operator $s_i$ such that
\beq\label{big simplex}
\tr{s_i s_j}=\vec{s}_i\cdot\vec{s}_j.
\eeq
The $\{\vec{s}_i\}$ are the operator representation of the simplex, and in particular $\sum_i s_i{=}0$. The $d^2$ trace-one operators,
\beq\label{sic op}
\sigma_i=\frac1{d}(1+s_i),
\eeq
respect the symmetry of the simplex, 
\beq\label{tr sic op}
\tr{\sigma_i\sigma_j}=\left\{
  \begin{array}{ll}
    \frac1{d^2}(d+\al) & \quad \text{for}\; i=j\\
    \frac1{d^2}\Bigl(d-\al\frac{1}{d^2-1}\Bigr) & \quad \text{for}\; i\neq j,
  \end{array} \right.
\eeq
but with $\sum_i \sigma_i{=}d$. Thus, a $(d^2{-}1)$-simplex in ${\mathbb R}^{d^2{-}1}$ corresponds to $d^2$ symmetric, trace-one, hermitian operators acting on ${\cal H}^d$, and sum up to the $d$ times the identity. We simply call them symmetric operators (SO). 

If the $\sigma$s are non-negative, then the set $\{\frac1{d}\sigma_i\}$ form a SIC POM. In the particular case of  $\al{=}d(d{-}1)$, 
\beq
\tr{\sigma_i\sigma_j}=\left\{
  \begin{array}{ll}
    1 & \quad \text{for}\; i=j\\
    \frac1{d+1} & \quad \text{for}\; i\neq j.
  \end{array} \right.
\eeq
If for this value of $\al$ the $\sigma$s are non-negative  then they are also of rank-one, that is projectors onto kets (pure states) in ${\cal H}^d$. In this case $\{\frac1{d}\sigma_i\}$ is a rank-one SIC POM.  Actually, if $\sigma_i{<}0$ one can rescale it by the magnitude of its smallest eigenvalue,  rendering the corresponding operator $\sigma_i$ positive semi-definite. Since regular simplex exists in any finite-dimensional real vector space, a (high-rank) SIC POM also exist in any finite-dimensional Hilbert space.

\subsection{Mutually unbiased hermitian operators}\label{simplex sub b}
Next, we consider the following collection of traceless hermitian operators, $t_m^{(b)}$, with $b{=}0,\ldots,d$ and $m{=}0,\ldots ,d{-}1$, such that,
\beq\label{simplex t}
\tr{t_m^{(b)}t_{m'}^{(b')}}=\left\{
  \begin{array}{ll}
    0 & \quad \text{for}\; b\neq b'\\
    \beta & \quad \text{for}\;b=b'\;\text{and}\; m=m'\\
    \beta \bigl(-\frac1{d-1}\bigr)& \quad \text{for}\;b=b'\;\text{and}\; m\neq m'.
  \end{array} \right.
\eeq
Since the trace operation corresponds to the scalar product in a real vector space, this equation has two implications: (a) operators with different $b$ label belong to orthogonal subspaces in ${\mathbb R}^{d^2{-}1}$, and (b) the $d$ operators $\{t_m^{(b)}:m{=}0,\ldots,d{-}1\}$ form a regular simplex in a $(d{-}1)$-dimensional subspace of ${\mathbb R}^{d^2{-}1}$, that is a $(d{-}1)$-simplex. Since there are at most $d{+}1$ orthogonal subspaces ${\mathbb R}^{d{-}1}$ in ${\mathbb R}^{d^2{-}1}$, there are at most $d{+}1$ such orthogonal $(d{-}1)$-simplexes. The maximum set of regular simplexes exists in any  finite dimension inasmuch as for any finite $d$, ${\mathbb R}^{d^2{-}1}$ can be decomposed into $d{+}1$ orthogonal subspaces. 

The trace-one hermitian operators, $\tau_m^{(b)}$, 
\beq\label{mu op}
\tau_m^{(b)}=\frac1{d}(1+t_m^{(b)}),
\eeq
inherit the symmetric structure of  $t_m^{(b)}$,
\beq\label{tr mu op}
\tr{\tau_m^{(b)}\tau_{m'}^{(b')}}=\left\{
  \begin{array}{ll}
    \frac1{d} & \quad \text{for}\; b\neq b'\\
    \frac1{d^2}\bigl(d+\beta\bigr) & \quad \text{for}\;b=b'\;\text{and}\; m=m'\\
    \frac1{d^2}\Bigl(d-\beta\frac{1}{d-1}\Bigr)& \quad \text{for}\;b=b'\;\text{and}\; m\neq m'.
  \end{array} \right.
\eeq
We note that ${\sum_m}t_m^{(b)}{=}0$, and, consequently, ${\sum_m}\tau_m^{(b)}{=}1$ $\forall b$.

Evidently, the overlap of the any two operators belonging to different sets (labeled by $b$) is a constant, $1/d$, that is the operators with different $b$ label are mutually unbiased (MU). Consequently, in any finite-dimensional Hilbert space one can find $d{+}1$ sets of MU operators (MUO). In general the $\tau_m^{(b)}$s do not represent statistical operators, but when they do, each set of operators $\{\tau_m^{(b)}:m{=}0,\ldots,d{-}1\}$ form a POM. The $d{+}1$ POMs are MU, that is they form a complete set of MU POM. Beyond the mathematical property of Eq.~(\ref{tr mu op}), the unbiasedness has a physical meaning: If a system is prepared in a state $\tau_m^{(b)}$, then the transition probability to any state $\tau_m^{(b')}$ with $b{\neq}b'$ is independent of the chosen states.
The $d{+}1$ MU POM are are informationally complete. Like for the SO case, if $\tau_m^{(b)}{<}0$ one can
rescale it by the magnitude of its smallest eigenvalue, rendering it positive semi-definite. 

For value $\beta=d(d-1)$, 
\beq
\tr{\tau_m^{(b)}\tau_{m'}^{(b')}}=\left\{
  \begin{array}{ll}
    \frac1{d} & \quad \text{for}\; b\neq b'\\
    1 & \quad \text{for}\;b=b'\;\text{and}\; m=m'\\
    0& \quad \text{for}\;b=b'\;\text{and}\; m\neq m'.
  \end{array} \right.
\eeq
If the $\tau$s are also non-negative for this value of $\beta$ then they are necessarily rank-one projectors onto MUB, cf. Eq.~(\ref{mub def}).

\section{Geometrical relation between MUO and SO}\label{geometry}
In this section we show that MUO and SO have the same relation as points and lines in specific type if finite projective plane geometries. We study the implication of this structure in general and for the two special cases of MUB and SIC POMs.
\subsection{Brief review of finite projective plane geometries}\label{geometry sub a}
We briefly review some of the properties of finite projective planes that are of relevance for our discussion \cite{wootters06,revzen11,planebook}. A finite projective plane is a geometrical structure consists of finite number of points and lines such that,
\begin{itemize}
	\item any two distinct lines intersect in one and only one point, 
 \item any two distinct points are connected by one and only one line.
\end{itemize}

In a nutshell, a finite affine plane geometry (APG) is a finite projective plane geometry with additional properties. It is composed of $d^2$ points and $d(d{+}1)$ lines, such that
\begin{itemize}
	\item each line contains $d$ points,
	\item every point is contained in $d{+}1$ lines,  
	\item Given any line L, and any point P not on L, there is exactly one line containing P that does not intersect L.
\end{itemize}
Accordingly, the lines could be grouped into $d{+}1$ groups, each contains $d$ parallel (i.e., not intersecting) lines, and any two lines each from different sets intersect in one and only one point. The number $d$ is called the order of the APG. It can be shown that APG could be constructed in prime-power orders. 

\begin{figure}[t]
\centering
\includegraphics[scale=1.]{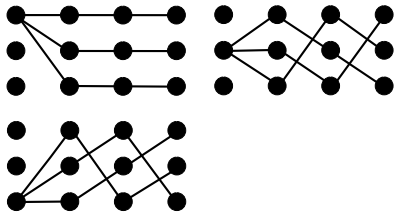}
\caption{ADPG of order 3. The 12 points are grouped into 4 sets (columns) of 3 points. The points in a column are not connected by any line. There are 9 lines in the DAPG of order 3. Each line is define uniquely by two points. An example of the possible 9 lines is sketched.}
\label{fig:dim3}
\end{figure}

The main analysis in this section concerns with dual APGs (DAPGs). A DAPG can be constructed from an APG by interchanging points with lines. Therefore, a DAPG of a prime-power order $d$ is composed of $d(d{+}1)$ points and $d^2$ lines, such that
\begin{itemize}
	\item every point is contained in $d$ lines
	\item every line contains $d{+}1$ points,  
	\item Given any point P, and any line L not containing P, there is exactly one point on line L that does not connected to P.
\end{itemize}
Since any two distinct lines intersect in one and only one point (according to the features of projective planes), these properties imply that the $d(d{+}1)$ points can be grouped into $d{+}1$ groups, each of $d$ points, such that the points in each group are not connected by any line. This property is the dual property of parallel lines in APG. Therefore, one can visualize the DAPG by a grid of $d(d{+}1)$ points arranged in $d{+}1$ columns of $d$ points, such that, points on each column are not connected by any line, and any point on a column is connected to any other point (not on its column) by $d$ lines. The points are now labeled by two coordinates $(m,b)$ which indicates the raw and column `position' of the point on the grid. As an illustrative example let us look at the (almost) simplest case, the DAPG of order $d{=}3$. It contains 12 points that are connected by 9 lines, with 4 points per line. The lines intersect in one and only one point. The points can be grouped into 4 sets of 3 points,  with points at a set not connected by any line, see Fig~\eqref{fig:dim3}.

\subsection{The geometrical relation}\label{geometry sub b}
In the case of a prime-power order, a geometrical relation between the $d^2$ SO of Eq.~\eqref{sic op} and the $d(d{+}1)$ MUO of Eq.~\eqref{mu op} can be formulated in two possible ways. Either by identifying the $d^2$ points of the APG with the $d^2$ traceless hermitian operators $s_i$, or by affiliating the points  $(m,b)$ of the DAPG with $t_m^{(b)}$. In what follows we will study the second possibility for which, the $d^2$ lines constructed from the $d(d{+}1)$ points in the DAPG correspond to the traceless hermitian operators $s_i$s. For completeness we remark that, in the first possibility the $d(d{+}1)$ lines constructed from the $d^2$ points in the APG correspond to the traceless hermitian operators $t_m^{(b)}$s, in a way that operators with the same $b$ label correspond to parallel lines.

Consider the second identification. We start by associating the $d(d{+}1)$ points on the DAPG grid, with the hermitian traceless operators $t_m^{(b)}$ of Eq.~\eqref{simplex t} with $b{=}0,\ldots,d$, and $m{=}0,1,\ldots,d{-}1$. The correspondence is such that the traceless operator $t_m^{(b)}$ is associated with the point $(m,b)$ on the grid. In this context we call the $t_m^{(b)}$s (traceless) point operators. Depending on the context, and where there is no ambiguity, we shall refer to $\tau_m^{(b)}$ as (trace-one) point operator as well. We recall that the $t_m^{(b)}$s form $d{+}1$ orthogonal $(d{-}1)$-simplexes in ${\mathbb R}^{d^2{-}1}$, where $b$ labels the orthogonal subspace, and that the corresponding  $\tau_m^{(b)}$s form $d{+}1$ sets of MUO, each set is labeled by $b$.

Next, we identify the operators that are associated with lines in the DAPG---the line operators. There are $d^2$ lines in this plane, each contains $d$ points, one point from each column. The lines intersect in one and only one point. We label the lines by $\mu$, $\mu{=}0,\ldots,d^2{-}1$. We use the notation $(m,b){\in}\mu$ to denote points $(m,b)$ on the line $\mu$, and $\mu{\ni}(m,b)$ to denote lines that go through the point $(m,b)$.  We identify the line operator, $l_\mu$, with the sum of the (traceless) point operators $t_m^{(b)}$ that correspond to the points on the $\mu$th line,
\beq\label{line op}
l_\mu=\sum_{(m,b)\in\mu}{t_m^{(b)}}.
\eeq
By construction, the $l$s are hermitian traceless operators, and therefore they are elements of ${\mathbb R}^{d^2{-}1}$. Actually, the $l$s form a $(d^2{-}1)$-simplex in ${\mathbb R}^{d^2{-}1}$, as from Eq.~(\ref{simplex t}) we obtain
\begin{align}\label{line simplex}
\tr{l_\mu^2}&=\sum_{(m,b)\in\mu}\tr{t_m^{(b)}t_m^{(b)}}=\beta(d+1),\nn\\
\tr{l_\mu l_{\mu'}}&=\sum_{(m',b')\in\mu'}\sum_{(m,b)\in\mu}\tr{t_{m'}^{(b')}t_m^{(b)}}=\beta\Bigl(1-\frac{d}{d-1}\Bigr)=\beta(d+1)\Bigl(-\frac{1}{d^2-1}\Bigr),\;\text{for}\;\mu\neq\mu',
\end{align}
cf. Eq~\eqref{equi vec} with $\al{=}\beta(d{+}1)$. We note that $\sum_m t_m^{(b)}=\sum_\mu l_\mu=0$.

The point operators can be re-written in terms of the line operators as
\beq\label{point op}
t_m^{(b)}=\frac1{d}\sum_{\mu\ni(m,b)}{l_\mu}.
\eeq
This equation is derived from Eq.~(\ref{line op}) by observing that $d$ lines passing  through a point, and a point is connected to all other points not in its column through those lines. With this association, the inner product between a point operator and a line operator is given by
\beq\label{point line ip}
\tr{t_m^{(b)}l_\mu}=\left\{
  \begin{array}{ll}
    \beta & \quad \text{if }(m,b)\in\mu\\
    \beta \bigl(-\frac1{d-1}\bigr)& \quad \text{if }(m,b)\notin\mu.
  \end{array} \right.
\eeq

Now,  trace-one point, and line, operators can be constructed from the traceless operators. The trace-one (hermitian) point operators  $\tau_m^{(b)}$ are given in Eq.~(\ref{mu op}), and the trace-one (hermitian) line operators are given by,
\beq\label{line to sic}
\lambda_\mu=\frac{1}{d}(1+l_\mu),\; \mu=0,\ldots,d^2-1.
\eeq
We can also write the  line operators $\lambda_\mu$ in terms of the point operators $\tau_m^{(b)}$ and vice versa as,
\beq\label{lambda in tau}
\lambda_\mu=\frac1{d}\bigl(1+\sum_{(m,b)\in\mu}\!\!\!t_m^{(b)}\bigr)=\frac1{d}\bigl(1+\!\!\!\sum_{(m,b)\in\mu}\!\!\!(d\tau_m^{(b)}-1)\bigr)=\frac1{d}\bigl(1+d\!\!\!\sum_{(m,b)\in\mu}\!\!\!\tau_m^{(b)}-(d+1)\bigr)=\!\!\!\sum_{(m,b)\in\mu}\!\!\!\tau_m^{(b)}-1,
\eeq
and, similarly,
\beq\label{tau in lambda}
\tau_m^{(b)}=\frac1{d}\!\!\!\sum_{\mu\ni(m,b)}\lambda_\mu.
\eeq
Indeed the $\lambda_\mu$s are SO as they have equal pair-wise inner product,
\beq\label{line sic op}
\tr{\lambda_\mu\lambda_{\mu'}}=\left\{
  \begin{array}{ll}
    \frac1{d^2}(d+\beta(d+1)) & \quad \text{for}\; \mu=\mu'\\
    \frac1{d^2}\Bigl(d-\beta\frac{1}{d-1}\Bigr) & \quad \text{for}\; \mu\neq\mu'.
  \end{array} \right.
\eeq
Following Eq.~(\ref{point line ip}), the inner product between a line operator and a point operator is given by
\beq\label{tr1 point line ip}
\tr{\tau_m^{(b)}\lambda_\mu}=\left\{
  \begin{array}{ll}
    \frac1{d^2}(d+\beta) & \quad \text{if }(m,b)\in\mu\\
     \frac1{d^2}\bigl(d+\frac\beta{d-1}\bigr)& \quad \text{if }(m,b)\notin\mu.
  \end{array} \right.
\eeq
The above relations suggest two particular interesting values of $\beta$, $\beta{=}d(d{-}1)$ and $\beta{=}d(d{-}1)/(d{+}1)$, for which we devote the next two subsections.

\subsection{Case study: $\beta{=}d(d{-}1)$}\label{geometry sub c}
For the value $\beta{=}d(d{-}1)$, the point operators satisfy
\beq
\tr{\tau_m^{(b)}\tau_{m'}^{(b')}}=\left\{
  \begin{array}{ll}
    \frac1{d} & \quad \text{for}\; b\neq b'\\
    1& \quad \text{for}\;b=b'\;\text{and}\; m=m'\\
    0& \quad \text{for}\;b=b'\;\text{and}\; m\neq m',
  \end{array} \right.
\eeq
If for this value of $\beta$ the point operators are non-negative, then they are necessarily of rank-one. This is the case where the point operators are projectors onto $d{+}1$ MUB. 

Indeed for prime-power $d$, there exist $d{+}1$ MUB such that Eq.~(\ref{mub def}) holds, and the point operators are given by $\tau_m^{(b)}=|m;b\rangle\!\langle m;b|$ with $b=0,1,\ldots,d$ and $m=0,1,\ldots,d-1$. 

For the same value of $\beta$, the line operators constructed from the above point operators satisfy
\beq
\tr{\lambda_\mu\lambda_{\mu'}}=\left\{
  \begin{array}{ll}
    d & \quad \text{for}\; \mu=\mu'\\
    0& \quad \text{for}\; \mu\neq\mu'.
  \end{array} \right.
\eeq
Therefore, in this case the $\lambda_\mu$s are linearly independent and form a basis for the space of hermitian operators. By using  Eq.~(\ref{lambda in tau}) we can now write the operator basis in terms of the projectors on MUB,
\beq
\lambda_\mu=\!\!\!\sum_{(m,b)\in\mu}\!\!\!|m;b\rangle\!\langle m;b|-1.
\eeq
It is interesting to note that the inner product of Eq.~(\ref{point line ip}) between the line operators and the point operators, follow the geometrical intuition,
\beq
\tr{\tau_m^{(b)}\lambda_\mu}=\left\{
  \begin{array}{ll}
    1 & \quad \text{if }(m,b)\in\mu\\
    0& \quad \text{if }(m,b)\notin\mu,
  \end{array} \right.
\eeq
that is, it is equals to one if the point is on the line, or else it is zero.

\subsection{Case study: $\beta{=}d(d{-}1)/(d{+}1)$}\label{geometry sub d}
Another interesting value for $\beta$ is $\beta{=}d(d{-}1)/(d{+}1)$, for which 
\beq\label{line beta2}
\tr{\lambda_\mu\lambda_{\mu'}}=\left\{
  \begin{array}{ll}
    1 & \quad \text{for}\; \mu=\mu'\\
    \frac1{d+1}& \quad \text{for}\; \mu\neq\mu'.
  \end{array} \right.
\eeq
In this case the line operators $\lambda_\mu$ are  geometrically constructed from point operators with
\beq\label{point beta2}
\tr{\tau_m^{(b)}\tau_{m'}^{(b')}}=\left\{
  \begin{array}{ll}
    \frac1{d} & \quad \text{for}\; b\neq b'\\
    \frac2{d+1}& \quad \text{for}\;b=b'\;\text{and}\; m=m'\\
    \frac1{d+1}& \quad \text{for}\;b=b'\;\text{and}\; m\neq m'.
  \end{array} \right.
\eeq
If the $\lambda$s are non-negative for this value of $\beta$ then they are also of rank-one, and actually they are (rank-one) projectors onto SIC-states, cf. Eq.~(\ref{sic}). The general structure of the SIC-states in finite-dimensional Hilbert space is not known. In fact only few SIC-states are known in specific dimensions, see e.g., Ref~\cite{scott10}. Let us assume the existence of SIC-states in some prime-power dimension, $d$. Then we can write the line operators of Eq.~(\ref{line beta2}) as $\lambda_\mu=\ket{\mu}\!\bra{\mu}$. The operators that form the SIC POM associated with $\ket{\mu}$ are $\frac1{d}\lambda_\mu$. 

According to Eq.~(\ref{tau in lambda}), the line point operators are given by
\beq\label{point in line beta2}
\tau_m^{(b)}=\frac1{d}\sum_{\mu\ni(m,b)}\ket{\mu}\!\bra{\mu},
\eeq
and therefore they can be regarded as statistical operators. Moreover, since for any $b$ value $\sum_m\tau_m^{(b)}{=}1$, the set $\{\tau_m^{(b)}\}_{m{=}0}^{d{-}1}$ form a POM, and the set of all $\tau_m^{(b)}$ forms an IC POM. Following the discussion in Sec.~\ref{simplex} and as can  be seen from Eq.~(\ref{point beta2}), the  point operators $\tau_m^{(b)}$s for different values of $b$ are MU, hence the sets $\{\tau_m^{(b)}\}_{m{=}0}^{d{-}1}$ with different $b$ label form $d{+}1$ MU POM. The only possibility to construct $d{+}1$ sets of MU POM is by constructing the POMs from operators which are diagonal in the $d{+}1$ sets of MUB. To be more explicit, to construct $d{+}1$ sets of MU POM, the probability-operators $\tau_m^{(b)}$ must be diagonal in the basis $\{|m;b\rangle\}$,
\beq
\tau_m^{(b)}=\sum_{k=0}^{d-1}\ket{k;b}(p_{m}^{(b)})_{k}\bra{k;b},\; {\rm with}\;\sum_{k=0}^{d-1}(p_m^{(b)})_{k}=1,
\eeq
so that the unbiasedness property is satisfied by construction: 
\beq
\tr{\tau_m^{(b)}\tau_{m'}^{(b')}}=\frac1{d}\sum_{k,k'}(p_{m}^{(b)})_{k}(p_{m'}^{(b')})_{k'} =\frac1{d},\; \forall\; b\neq b'.
\eeq
We denote by $\vec{p}_{m}^{(b)}$ the `probability vector' whose coordinates are the probabilities $(p_{m}^{(b)})_{k}$. The inner product of two probability-operators of the same basis yields non-trivial conditions that the probabilities $(p_{m}^{(b)})_{k}$ should satisfy,
\beq\label{pm cond}
\tr{\tau_m^{(b)}\tau_{m'}^{(b)}}=\sum_{k}(p_{m}^{(b)})_{k}(p_{m'}^{(b)})_{k}=\left\{
  \begin{array}{ll}
    \frac2{d+1}& \quad \text{for}\; m=m'\\
    \frac1{d+1}& \quad \text{for}\; m\neq m'.
  \end{array} \right.
\eeq
This result was also obtained in Ref.~\cite{appleby09}, but in a slightly different context. In general the spectrum of each probability-operator $\tau_m^{(b)}$, $\{(p_{m}^{(b)})_{k}:k{=}0,\ldots,d{-}1\} $, is different. However some conditions on the spectrum of $\tau_m^{(b)}$ with the same value $b$, can be derived as follows. We note that $\sum_{k}(p_m^{(b)})_{k}{=}1$ and $\sum_m\tau_m^{(b)}{=}1$; the latter equation means that for every $k$ value, $\sum_m(p_{m}^{(b)})_{k}{=}1$. A possible solution for this system of equations (under the constraint that the $p$s are probabilities), is that the for a given $b$ there is only one probability vector $\vec{p}_{0}^{(b)}$ and all other probability vectors $\vec{p}_{m}^{(b)}$ are related to it by some permutation, which without loss of generality we can take as the cyclic permutation, $\vec{p}_{m}^{(b)}{=}C_m\vec{p}_{0}^{(b)}$. Indeed, if for a given $b$ we construct a matrix whose raws are the probability vectors $\vec{p}_{m}^{(b)}$, then the sums $\sum_{k}(p_m^{(b)})_{k}{=}1$ and $\sum_m(p_{m}^{(b)})_{k}{=}1$ imply that the sum of each raw and column of that matrix is one. Though a possible solution is that the raws and columns are permutation of a `seed' probability vector, this is not the only possible solution. 

As we shall see in the Appendix that SIC-states which are generated by the action of the HW group elements induce the property that operators $\tau_m^{(b)}$ with the same $b$ label have the same spectrum (while those with different $b$ label do not necessarily have the same spectrum). Actually, in the Appendix we analyze the structure of some known SIC-states from the point of view of the spectrum of the `underlying' MU POM. We find that MU POM that correspond to {\it different} $b$ label share the {\it same} spectrum. As we discuss in the Appendix, this is somewhat surprising since it hints on a symmetry of the SIC-states that go beyond th HW group symmetry. 

Let us therefore consider the solution  where $\vec{p}_{m}^{(b)}{=}C_m\vec{p}_{0}^{(b)}$. In this case, the conditions of Eq.~(\ref{pm cond}) are further simplify,
\beq\label{pm cond simple}
\sum_{k}(p_{0}^{(b)})_{k}(p_{0}^{(b)})_{k\oplus m}=\left\{
  \begin{array}{ll}
    \frac2{d+1}& \quad \text{for}\; m=0\\
    \frac1{d+1}& \quad \text{for}\; m=1,\ldots,\frac{d-1}{2},
  \end{array} \right.
\eeq
where $\oplus$ stands for plus modulo $d$, and because the modular structure of this equations, the set of equations with  $m{=}1,\ldots,\frac{d-1}{2}$ is equal to the set of equation with $m{=}\frac{d-1}{2}+1,\ldots,d-1$. We note in passing that summing all these equations for $m{=}0,\ldots,d-1$, we obtain the trivial equation, $(\sum_{k}(p_{0}^{(b)})_{k})^2=1$.

As mentioned before, this structure $\vec{p}_{m}^{(b)}{=}C_m\vec{p}_{0}^{(b)}$ is rooted in the structure of the SIC-states. To see this, we express the point operators in terms of the line operators (which are projectors onto the SIC-states), Eq.~(\ref{point in line beta2}), with accordance to the geometry relation between points and lines in the DAPG. The set of points that are not connected by a line correspond to point operators with the same $b$ label (these are the point located on the same column of grid of points visualizing the DAPG, see e.g. Fig.~\ref{fig:dim3}). On the other hand $d$ lines going through a point, and therefore the the total number of lines, $d^2$, go through the $d$ points on a column, that is, with the same $b$ label. This means that for a every given column there is a unique grouping of the lines into $d$ sets each set contain $d$ lines that go through a one and only one point on the column (as can be seen in Fig.~\ref{fig:dim3}). If the point operators with the same $b$ label have the same spectrum, then the sums $\sum_{\mu\ni(m,b)}\ket{\mu}\!\bra{\mu}$ and $\sum_{\mu\ni(m',b)}\ket{\mu}\!\bra{\mu}$ are related to each other by  a unitary transformation. This transformation is the one that cyclic shifts the states in the basis $\{|m;b\rangle\}$. This impose constraints on the possible structure of the group that generate a set of SIC-states from a fiducial state. More on this matter for the special case of HW SIC POM in the Appendix, where we show that for that case $\vec{p}_{m}^{(b)}{=}C_m\vec{p}_{0}^{(b)}$.

The geometrical relation between MU POM (points) and projectors onto SIC-states (lines) may provide us tools to study the structure of SIC-states in particular dimensions. Such analysis could provide us with new insights into  possible structure of SIC-states in prime-power dimensions. From Eq.~(\ref{lambda in tau}) and from the above discussion we realize that in prime-power  dimensions, projectors onto SIC-states can be written as sums of operators that form MU POM (minus the identity), where each operator in the sum is diagonal in a given MUB, and no pair of operators in a sum is diagonal in the same basis. This could be used to construct SIC-states with HW group covariance structure. In this case, it is enough (though it might not be a simple task) to find the fiducial line operator whose structure, without loss of generality, is given by
\begin{equation}\label{fid line}
\ket{\mu_{0}}\!\bra{\mu_{0}}=\sum_b\tau_0^{(b)}-1.
\end{equation}
It is therefore equivalently to find probability-operators, $\tau_0^{(b)}$, each diagonal in the MUB labeled by $b$, such that the diagonal elements satisfy Eq.~(\ref{pm cond simple}) and the eigenvalues of their sum is either 2 or 1, the latter with multiplicity $d{-}1$. 

We examplify the above considerations for the case of $d=2$ (qubit) and $d=3$ (qutrit). First, we wish to construct a SIC POM for a qubit from MU POM. We construct the latter from the three eigenbases of the Pauli operators, $\sigma_x,\sigma_y$, and $\sigma_z$ for a qubit. We should construct three groups of two point operators (six operators in total), such that each two are diagonal in one of these bases,  $\{\ket{0_a},\ket{1_a}\}$ with $a=x,y,z$. In particular, we take $\tau_0^{(a)}=\ket{0_a}p_0^{(a)}\bra{0_a}+\ket{1_a}(1-p_0^{(a)})\bra{1_a}$ and $\tau_1^{(a)}=\ket{0_a}(1-p_0^{(a)})\bra{0_a}+\ket{1_a}p_0^{(a)}\bra{1_a}$, where the $p$s must abide by Eq.~(\ref{pm cond simple}),
\begin{align}\label{p cond qubit}
(p_{0}^{(a)})^2+(1-p_{0}^{(a)})^2&=\frac2{3}\nonumber\\
2p_{0}^{(a)}(1-p_{0}^{(a)})&=\frac1{3}.
\end{align}
There only one (non-trivial)  solution for this equation is $p_{0}^{(a)}{\equiv}p=\frac1{6}(3+\sqrt3)$ (the other solution is $1-p_{0}^{(a)}$), independently of the basis label $a$. Now, if for this $p$ value the operator,
\begin{equation}
\lambda_{0}=\sum_{a=x,y,z}\tau_0^{(a)}-1,
\end{equation}
is a rank-one projector, then, according to the discussion above, $\lambda_{0}$ is a fiducial projector, from which we can generate rank-one SIC POM by the action of the Pauli operators. Indeed the operator
\beq
\lambda_{0}=\sum_{a=x,y,z}\Bigr(\ket{0_a}\frac1{6}(3+\sqrt3)\bra{0_a}+\ket{1_a}\frac1{6}(3-\sqrt3)\bra{1_a}\Bigl)-1=\frac1{2}\bigl(1+\frac{1}{\sqrt3}\sum_{a=x,y,z}\sigma_a\bigr),
\eeq
is a rank-one projector, and it can be used to generated a SIC POM for a qubit by the action of Pauli operators.

Next, we consider the qutrit case, $d=3$. For prime dimension there are  $d+1$ MUB that could be defined following Eq.~(\ref{mub}). We consider these bases for the qutrit, and construct the point operators 
\beq
\tau_0^{(b)}=\ket{0;b}p_0^{(b)}\bra{0;b}+\ket{1;b}p_1^{(b)}\!\bra{1;b}+\ket{2;b}p_2^{(b)}\!\bra{2;b},
\eeq
where the $p$s satisfy the equations,
\begin{align}\label{p cond qutrit}
1-p_{0}^{(b)}-p_{1}^{(b)}&=p_{2}^{(b)},\nonumber\\
(p_{0}^{(b)})^2+(p_{1}^{(b)})^2+(p_{2}^{(b)})^2&=\frac1{2},\nonumber\\
p_{0}^{(b)}p_{1}^{(b)}+p_{1}^{(b)}p_{2}^{(b)}+p_{2}^{(b)}p_{0}^{(b)}&=\frac1{4}.
\end{align}
The solution for this system of equations (up to relabeling of the $p$s) is 
\beq
p_{0}^{(b)}=\frac{1}{2} \Bigl(1-p_{1}^{(b)}+\sqrt{2 p_{1}^{(b)}-3 (p_{1}^{(b)})^2}\Bigr),
\eeq
with the constraints $0\leq p_{0,1,2}^{(b)}\leq1$. If we take the probabilities to be independent of the basis label, $p_0=p_1=1/2$ and $p_2=0$, then the operator
\begin{equation}
\lambda_{0}=\frac1{2}\sum_{b=0}^3\bigl(\ket{0;b}\!\bra{0;b}+\ket{1;b}\!\bra{1;b}\bigr)-1,
\end{equation}
is a rank-one projector onto the ket $\frac1{\sqrt2}(\ket{0}-\omega^2\ket{1})$, which is a fiducial vector for SIC-states by the action of the HW group element in dimension three.

\subsection{Discrete `phase-space' quasi distribution}\label{geometry sub e}
In the previous subsections, we have identify the $d(d{+}1)$ points of a DAPG with the traceless hermitian operators $t_m^{(b)}$  which form $d{+}1$ orthogonal $(d{-}1)$-simplexes in ${\mathbb R}^{d^2{-}1}$. The trace-one operators constructed from $t_m^{(b)}$, $\tau_m^{(b)}$, of Eq.~\eqref{mu op} form $d{+}1$ sets of MUO acting on ${\cal H}^d$. The $d^2$ lines of the DAPG, were then identified as sums of the corresponding point operators, resulting in a traceless hermitian operators $l_\mu$. The line operator form a $(d^2{-}1)$-simplex in ${\mathbb R}^{d^2{-}1}$, and the trace-one operators of Eq.~\eqref{line to sic} constructed from $l_\mu$ are $d^2$ SO.

Thus far, we have identify the MUO $\tau_m^{(b)}$ with points in a DAPG (when the order $d$ is prime-power) and the lines corresponds to SO. If we interchange the role of points and lines we end up with identifying SO as points in the APG and MUO with lines. This association is related to the question of existence of rank-one SIC POMs and MUB (at least for prime-power dimensions). The question of the existence of rank-one SIC POMs is known to be very difficult to answer, while the existence of MUB in prime-power dimension was proven. In this geometrical context, it means that it is relatively easy to find $d^2$ SO (not necessarily positive semi-definite) from which one can construct (by the machinery of points and lines in an APG or DAPG) projectors onto $d{+}1$ MUB. Such a construction is given by Wootters for prime dimension \cite{wootters87}. However it is notoriously hard to find rank-one SIC POMs and therefore apparently it is a hard problem to find $d{+}1$ sets of MUO (being negative and high-rank in general) from which one can construct a rank-one SIC POM.

With these associations of points and lines to either MUO or SO, the  discrete version of a phase-space quasi-distribution function is called for. 
Suppose that the lines in either geometries correspond to (trace-one) non-negative operators, i.e., either to MU POM or to (high-rank) SIC POM. The operators correspond to the underlying points may be negative. In this case, the trace of a positive semi-definite trace-one operator (that is a statistical operator) with a point operator can be considered as a discrete version of a phase-space quasi-distribution function. Since the sum of all the points operator corresponds to a line operator, then the sum of this phase-space function over points on a line is directly related to the probability of the corresponding POM element. Take for example the DAPG, and suppose that the $d^2$ lines correspond to a SIC POM elements $\lambda_\mu$ of Eq.~\eqref{line op} with $\lambda_\mu\geq0$ for all $\mu$ . The discrete quasi-probability distribution at point $(m,b)$ with $b{=}0,\ldots ,d$ and $m{=}0,\ldots ,d{-}1$, is given by
\beq
Q_{(m,b)}=\tr{\tau_m^{(b)}\rho},
\eeq
where $\rho$ is the statistical operator for the quantum system; and since the $\tau_m^{(b)}$s form MU POM, $Q_{(m,b)}{\geq}0$. Then by construction,
\beq
\frac{1}{d}\Bigl(\sum_{(i,k)\in\mu}Q_{(i,k)}-1\Bigr)=\tr{\frac{1}{d}\lambda_\mu\rho}=p_\mu,
\eeq
where $p_\mu$ is the probability of getting an outcome $\mu$ when measuring the SIC POM $\{\frac1{d}\lambda_\mu\}$.

\section{The role of HW group in DAPG}\label{hw-geometry}
This section concerns with the role of the Weyl-pair operators, $Z$ and $X$ defined in the Introduction, as generating lines in DAPG of prime orders. We show that if the points of a DAPG are associated with the operators $t_k^{(j)}$ of Eq.~\eqref{simplex t}, then the Weyl-pair generate all the operators associated with lines in this geometry [$l_\mu$ of Eq.~\eqref{line op}] by acting on an operator associated with a generic (fiducial) line.

\subsection{Generating SO from a fiducial operator}\label{hw-geometry sub a}
We start by parameterizing the  ($d{-}1$)-simplex $\{t_k^{(j)}:k{=}0,\ldots,d{-}1\}$ is regular simplex in ${\mathbb R}^{d{-}1}$. Consider a simplex in ${\mathbb R}^{d{-}1}$ (throughout this section $d$ is a prime number), whose Cartesian coordinates are given by the parametrization 
\beq\label{sim co}
(\vec{t}_k)_r=\Bigl(\cos\bigl(\frac{2\pi}{d}kr\bigr),\sin\bigl(\frac{2\pi}{d}kr\bigr)\Bigr),
\eeq
where $k$ labels the vector ($k{=}0,\ldots,d{-}1$), and $r$ labels pairs of coordinates ($r{=}1,\ldots,(d{-}1)/2$). With the above parametrization the set of vectors $\{\vec{t}_k:k{=}0,\ldots,d{-}1\}$ indeed form a regular simplex,
\begin{align}\label{sim in d-1}
\vec{t}_k\cdot\vec{t}_k&=\sum_{r=1}^{(d-1)/2}\Bigl[\cos^2\bigl(\frac{2\pi}{d}kr\bigr)+\sin^2\bigl(\frac{2\pi}{d}kr\bigr)\Bigr]=\frac{d-1}{2},\nn\\
\vec{t}_k\cdot\vec{t}_l&=\sum_{r=1}^{(d-1)/2}\Bigl[\cos\bigl(\frac{2\pi}{d}kr\bigr)\cos\bigl(\frac{2\pi}{d}lr\bigr)+\sin\bigl(\frac{2\pi}{d}kr\bigr)\sin\bigl(\frac{2\pi}{d}lr\bigr)\Bigr]\nn\\
&=\sum_{r=1}^{(d-1)/2}\cos\bigl(\frac{2\pi}{d}(k-l)r\bigr)=-\frac{1}{2}+\frac{\sin(\pi(k-l))}{2\sin\bigl(\frac{\pi}{d}(k-l)\bigr)}=-\frac{1}{2}\qquad\forall k\neq l,
\end{align}
c.f. with Eq.~\eqref{equi vec}. Note that in this parametrization, the vector $\vec{v}_0$ lies in $\frac{(d{-}1)}{2}$-dimensional subspace of ${\mathbb R}^{d{-}1}$, as half of its components are 1s and half of them are 0s, in an alternating fashion, $\vec{v}_0{=}(1,0,1,0,\ldots,1,0)$. The vector $\vec{t}_k\in{\mathbb R}^{d{-}1}$ is embedded in ${\mathbb R}^{d^2{-}1}$ corresponds to the operators $t_k^{(j)}$ where the $j$ labels the subspace in which $\vec{t}_k$ is embedded.

Next, consider the generators of the HW group, the Weyl-pair $Z$ and $X$, in prime dimension $d$, satisfying their defining properties
\begin{align}\label{hw gen}
Z^d&=X^d=1,\nn\\
\omega ZX&=XZ,
\end{align}
where $\omega{=}e^{\ii 2\pi/d}$ is the fundamental $d$th root of unity. The set $\{X^aZ^b:a,b{=}0,\ldots,d{-}1\}$ is pair-wise orthogonal, 
\beq
\tr{\Bigl(X^{a'}Z^{b'}\Bigr)^\dagger X^aZ^b}=d\delta_{a',a}\delta_{b',b},
\eeq 
and can be used as a basis of the $d^2$-dimensional Hilbert space of operators \cite{kibler08}. When the identity is excluded, the set $\{X^aZ^b:a,b{=}0,\ldots,d{-}1\}\backslash\{X^0Z^0\}$ form a basis for traceless operators acting on $d$-dimensional Hilbert space of kets.  In a prime dimension $d$,  $\{X^aZ^b:a,b{=}0,\ldots,d{-}1\}\backslash\{X^0Z^0\}$ can be divided into $d{+}1$ orthogonal subsets $\mathcal{V}$'s of $d{-}1$ commuting operators,
\begin{align}\label{sub groups}
&\mathcal{V}^{(d)}=\{Z^k,(Z^k)^\dagger:k=1,\ldots,(d-1)/2\},\nn\\
&\mathcal{V}^{(j)}=\{X^kZ^{kj},(X^kZ^{kj})^\dagger:k=1,\ldots,(d-1)/2\},\; j=0,\ldots,d{-}1.
\end{align}
Taking the following linear combination of operators within each subsets
\begin{align}\label{h op}
&h^{(d)}_{k}=\zeta_{d,k} Z^k + \zeta^*_{d,k}(Z^k)^\dagger,\nn\\
&g^{(d)}_{k}=-\ii\Bigl(\zeta_{d,k} Z^k - \zeta^*_{d,k}(Z^k)^\dagger\Bigr),\nn\\
&h^{(j)}_{k}=\zeta_{j,k} X^kZ^{kj} + \zeta^*_{j,k}(X^kZ^{kj})^\dagger,\nn\\
&g^{(j)}_{k}=-\ii\Bigl(\zeta_{j,k} X^kZ^{kj} - \zeta^*_{j,k}(X^kZ^{kj})^\dagger\Bigr),
\end{align} 
with complex numbers $\zeta$s, we obtain a basis for traceless hermitian operators acting on ${\cal H}^d$. This basis is an orthogonal basis as
\begin{align}
&\tr{h^{(j)}_{k}g^{(j')}_{k'}}=0,\nn\\
&\tr{h^{(j)}_{k}h^{(j')}_{k'}}=\tr{g^{(j)}_{k}g^{(j')}_{k'}}=2d|\zeta_{j,k}|^2\delta_{j,j'}\delta_{k,k'},
\end{align}
for $j{=}0,\ldots,d$ and $k{=}1,\ldots,(d{-}1)/2$.
The $d{+}1$ sets of the operators $\{h^{(j)}_{k},g^{(j)}_{k}:k=1,\ldots,(d-1)/2\}$, with $j{=}0,\ldots,d$ are mutually orthogonal and form a basis for the space of traceless hermitian operators acting on ${\cal H}^d$. Since the $d{+}1$ sets are mutually orthogonal, each span a $(d{-}1)$-dimensional subspace in ${\mathbb R}^{d^2{-}1}$. Note that we label these subspaces by an index taking over the values $j{=}0,\ldots,d$. Any traceless hermitian operator $t$,
\beq
t=\sum_{j=0}^{d}\sum_{k=1}^{\frac{d-1}{2}}\bigl(r_{j,k}h^{(j)}_{k}+s_{j,k}g^{(j)}_{k}\bigl),
\eeq
can be represented in this basis by a real vector $\vec{t}$ with $d^2{-}1$ components,
\beq
t\leftrightarrow\vec{t}=(r_{0,0},s_{0,0},\ldots,r_{0,\frac{d-1}{2}},s_{0,\frac{d-1}{2}},\ldots,r_{d,\frac{d-1}{2}},s_{d,\frac{d-1}{2}}).
\eeq

The action of the Weyl-pair operators $Z$ and $X$ on $h$ and $g$ is given by,
\begin{align}\label{z action on h}
&Z{\begin{pmatrix}h^{(d)}_{k}\\g^{(d)}_{k}\end{pmatrix}}Z^\dagger={\begin{pmatrix}h^{(d)}_{k}\\g^{(d)}_{k}\end{pmatrix}},\nn\\
&Z{\begin{pmatrix}h^{(j)}_{k}\\g^{(j)}_{k}\end{pmatrix}}Z^\dagger={\begin{pmatrix}\cos\bigl(\frac{2\pi}{d}k\bigr)&\sin\bigl(\frac{2\pi}{d}k\bigr)\\-\sin\bigl(\frac{2\pi}{d}k\bigr)&\cos\bigl(\frac{2\pi}{d}k\bigr)\end{pmatrix}}{\begin{pmatrix}h^{(j)}_{k}\\g^{(j)}_{k}\end{pmatrix}},
\end{align}
and
\begin{align}\label{x action on h}
&X^\dagger{\begin{pmatrix}h^{(d)}_{k}\\g^{(d)}_{k}\end{pmatrix}}X={\begin{pmatrix}\cos\bigl(\frac{2\pi}{d}k\bigr)&\sin\bigl(\frac{2\pi}{d}k\bigr)\\-\sin\bigl(\frac{2\pi}{d}k\bigr)&\cos\bigl(\frac{2\pi}{d}k\bigr)\end{pmatrix}}{\begin{pmatrix}h^{(d)}_{k}\\g^{(d)}_{k}\end{pmatrix}},\nn\\
&X^\dagger{\begin{pmatrix}h^{(j)}_{k}\\g^{(j)}_{k}\end{pmatrix}}X={\begin{pmatrix}\cos\bigl(\frac{2\pi}{d}kj\bigr)&\sin\bigl(\frac{2\pi}{d}kj\bigr)\\-\sin\bigl(\frac{2\pi}{d}kj\bigr)&\cos\bigl(\frac{2\pi}{d}kj\bigr)\end{pmatrix}}{\begin{pmatrix}h^{(j)}_{k}\\g^{(j)}_{k}\end{pmatrix}},
\end{align}
for $j=0,\ldots,d{-}1$. Therefore, the Weyl-pair act as rotation operators in two-dimensional real vector subspaces spanned by $h^{(j)}_{k}$ and $g^{(j)}_{k}$; see also~\cite{appleby09}. 

Consider the traceless hermitian operator
\beq\label{f op}
f^{(j)}_{0}=\sum_{k=1}^{\frac{d-1}{2}}h^{(j)}_{k},
\eeq
which is represented by a $(d{-}1)$-dimensional (reference) vector $\vec{f}^{(d)}_{0}{=}(1,0,1,0,\ldots,1,0)$ embedded in the $d$th (orthogonal) subspace of  ${\mathbb R}^{d^2{-}1}$. The action of $X^{a}$ on $f^{(d)}_{0}$ 
\beq
f^{(d)}_{a}\equiv X^{\dagger a}f^{(d)}_{0}X^{a}=\sum_{k=1}^{\frac{d-1}{2}}\Bigl[\cos\bigl(\frac{2\pi}{d}ka\bigr)h^{(d)}_{k}+\sin\bigl(\frac{2\pi}{d}ka\bigr)g^{(d)}_{k}\Bigr],
\eeq
result in an operator that is represented as a vector $\vec{f}^{(d)}_a$ in the subspace of ${\mathbb R}^{d^2{-}1}$ labeled by $d$ whose pair of components are given by
\beq
(\vec{f}^{(d)}_a)_k=\Bigl(\cos\bigl(\frac{2\pi}{d}ka\bigr),\sin\bigl(\frac{2\pi}{d}ka\bigr)\Bigr)
\eeq
with $k{=}1,\ldots,(d{-}1)/2$. The set of vectors $\{\vec{f}^{(d)}_a:a{=}0,\ldots,d{-}1\}$ form a regular simplex ($d$ equiangular vectors) in the [($d{-}1$)-dimensional] subspace of ${\mathbb R}^{d^2{-}1}$ labeled by $d$, c.f. Eq.~(\ref{sim co}) and~(\ref{sim in d-1}). The inner product of two corresponding operators in this simplex is given by
\begin{align}
\tr{f^{(d)}_{a}f^{(d)}_{b}}&=2d\sum_{k=1}^{\frac{d-1}{2}}|\zeta_{d,k}|^2\Bigl[\cos\bigl(\frac{2\pi}{d}ka\bigr)\cos\bigl(\frac{2\pi}{d}kb\bigr)+\sin\bigl(\frac{2\pi}{d}ka\bigr)\sin\bigl(\frac{2\pi}{d}kb\bigr)\Bigr]\nn\\
&=2d\sum_{k=1}^{\frac{d-1}{2}}|\zeta_{d,k}|^2\cos\bigl(\frac{2\pi}{d}k(a-b)\bigr),
\end{align}
which does not in general equal to the inner product of two vectors of the simplex. The requirement that the inner product of two operators equals to the inner product of the corresponding vectors of the simplex implies that
\beq\label{abs zeta}
|\zeta_{d,k}|^2=\frac1{2d}\quad\forall k=1,\ldots\frac{d-1}{2}.
\eeq

So far we have considered the action of $X^{a}$ on traceless hermitian operators of the form $f^{(d)}_{0}$. Now let us consider the action of  $X^{\dagger b}Z^{a}$ on  $f^{(j)}_{0}$ where $j\neq d$,
\beq
f^{(j)}_{ab}\equiv X^{\dagger b}Z^{a}f^{(j)}_{0}Z^{\dagger a}X^{b}=\sum_{k=1}^{\frac{d-1}{2}}\cos\bigl(\frac{2\pi}{d}k(a+jb)\bigr)h^{(j)}_{k}+\sin\bigl(\frac{2\pi}{d}k(a+jb)\bigr)g^{(j)}_{j,k}.
\eeq
Evidently $f^{(j)}_{ab}=f^{(d)}_{a{\oplus}jb}$ where $\oplus$ denotes addition modulo $d$, and therefore $f^{(j)}_{ab}$ as well corresponds to a regular simplex (in the $j$th orthogonal subspace of ${\mathbb R}^{d^2{-}1}$) where the trace of products of two operators equals to the inner product of the corresponding vectors of the simplex if $|\zeta_{j,k}|^2{=}\frac{1}{2d}$ for all $j{=}0,\ldots,d$ and $k{=}1,\ldots,(d{-}1)/2$. Importantly, $X^{\dagger b}Z^{a}$ rotates the reference vectors $\vec{f}^{(j)}_{0}$ to a another vector on the simplex in all but one subspaces  whose label $j$ satisfies $a{\oplus}jb{=}0$. We shall not fail to mention that, in accordance to the discussion in Sec.~\ref{simplex}, $f^{(j)}_{ab}$ the correspond to $d{+}1$ sets of MUO.

The (fiducial) traceless hermitian operator
\beq\label{ref line op}
l_{0}=\sum_{j=0}^{d}\sum_{k=1}^{\frac{d-1}{2}}h^{(j)}_{k}=\sum_{j=0}^{d}f^{(j)}_{0},
\eeq
is represented by the vector $\vec{l}_{0}{=}(1,0,1,0,\ldots,1,0)$ in  ${\mathbb R}^{d^2{-}1}$. Using the same letter to denote this operator as the one we used for the line operators in Sec.~\ref{geometry sub b} will be clear in what follows. The action of $X^{\dagger b}Z^{a}$ on $l_0$ corresponds to a rotation of vectors $\vec{f}^{(j)}_{0}$s on the regular simplexes at each of the $d{+}1$ orthogonal subspaces of ${\mathbb R}^{d^2{-}1}$,
\beq\label{lab}
l_{ab}\equiv X^{\dagger b}Z^{a}l_{0}Z^{\dagger a}X^{b}=f^{(d)}_{b}+\sum_{j=0}^{d-1}f^{(j)}_{ab}.
\eeq
For $|\zeta_{j,k}|^2{=}\frac{1}{2d}$ for all $j$ and $k$ the $l_{ab}$'s are $d^2$ SO,
\begin{align}
&\tr{l_{ab}^2}=(d+1)\frac{d-1}{2}\nn\\
&\tr{l_{ab}l_{a'b'}}=\frac{d-1}{2}+d\bigl(-\frac1{2}\bigr)=-\frac1{2}\quad\text{for}\;a\neq a',\;\text{and(or)}\;b\neq b'.
\end{align}
The last equation was obtained by realizing that for there is a unique solution for the equation $a{\oplus}jb{=}a'{\oplus}jb'$, for $a\neq a'$ and(or) $b\neq b'$. 

The above relations have clear geometrical interpretation. For $|\zeta_{j,k}|^2{=}\frac{1}{2d}$ for all $j$ and $k$, the operators $d(d{+}1)$ operators $f^{(d)}_{b}$ and $f^{(j)}_{ab}$ would correspond to the $d(d{+}1)$ points of the DAPG, while the $d^2$ operators $l_{ab}$ would corresponds to lines in this planes (these are the $l_\mu$ in Sec.~\ref{geometry sub b}). The operators associated with the points (point operators) and those associated with lines (lines operators) satisfy the axiomatic relation of points and lines of the DAPG. The operators $X^{\dagger b}Z^{a}$ generate the $d^2$ lines from a fiducial line operator $l_{0}$. As we discussed in this and in the previous section, the each group of $d$ point operators $f^{(d)}_{a{\oplus}jb}$ represent regular simplex in the $j$th subspace of ${\mathbb R}^{d^2{-}1}$, and therefore the  $d^2$ line operators (constructed from the points operators abiding by the axioms of the DAPG) correspond to a regular simplex in ${\mathbb R}^{d^2{-}1}$. If the operator $(1+l_0)/d$ is of rank-one, then $(1+l_{ab})/d^2$ would compose a rank-one SIC POM. 

\subsection{Condition for rank-one SIC POMs}\label{hw-geometry sub b}
The action of $X^{\dagger b}Z^{a}$ on the line operator of Eq.~(\ref{ref line op}) (with $|\zeta_{m,k}|^2{=}\frac{1}{2d}$ for all $k{=}1,\ldots,(d{-}1){/}2$) results in  $d^2$ traceless hermitian operators, $l_{\mu}$ (the pair of indices $ab$ are replaced by a single index $\mu$),  which correspond to a regular simplex in ${\mathbb R}^{d^2{-}1}$. This implies that the trace-one hermitian operators that constructed from $l_{\mu}$,
\beq
\sigma_{\mu}=\frac1{d}(1+\sqrt{\frac{2d}{d+1}}l_{\mu}),
\eeq
are SO, 
\begin{align}
&\tr{\sigma_{\mu}}=\tr{\sigma_{\mu}^2}=1,\nn\\
&\tr{\sigma_{\mu}\sigma_{\mu'}}=\frac1{d+1}\quad\text{for}\;\mu\neq\mu',
\end{align}
cf. Sec.~\ref{simplex}. If $\sigma_{0}$ is positive, then $\sigma_{\mu}$ are the elements of a rank-one SIC POM. 

Let us look in more details on the structure of $\sigma_{0}$. From Eqs.~(\ref{h op}), (\ref{f op}), and~(\ref{ref line op}) we get
\beq
\sigma_{0}=\frac1{d}\Bigl\{1+\frac1{\sqrt{d+1}}\sum_{k=1}^{\frac{d-1}{2}}\bigl[\bigl(e^{\ii\phi_{d,k}} Z^k + e^{-\ii\phi_{d,k}}(Z^k)^\dagger\bigr)+\sum_{j=0}^{d-1}\bigl(e^{\ii\phi_{j,k}} X^kZ^{kj} + e^{-\ii\phi_{j,k}}(X^kZ^{kj})^\dagger\bigr)\bigr]\Bigr\},
\eeq
The requirement that $\sigma_{0}$ is of rank-one is equivalent to the requirement that $\tr{\sigma_{0}^3}=1$. There are $(d^2{-}1)/2$ free phase-parameters to set in order to fulfill this requirement. Alternatively, if $\sigma_{0}$ is a  rank-one operator then $\sigma_{0}=\ket{\psi}\!\bra{\psi}$, where $\bra{n}\!\psi\rangle{=}a_ne^{\ii\varphi_n}$ for $n{=}0,\ldots,d{-}1$. Without loss of generality we take real and non-negative $a$s  and $\varphi_n{=}0$.

In a matrix representation the elements of $\sigma_{0}$ are given by
\begin{align}
&\bigl(\sigma_{0}\bigr)_{n,n}=\frac1{d}\Bigl[1+\frac1{\sqrt{d+1}}\sum_{k=1}^{\frac{d-1}{2}}\cos(\phi_{d,k}+\frac{2\pi}{d}kn)\Bigr],\nn\\
&\bigl(\sigma_{0}\bigr)_{n,n\oplus k}=\bigl(\sigma_{0}\bigr)_{n\oplus k,n}^*=\frac1{d\sqrt{d+1}}\sum_{j=0}^{d-1}e^{\ii\phi_{j,k}}\omega^{njk}
\end{align}
where  $n{=}0,\ldots,d{-}1$, $k{=}1,\ldots,(d{-}1)/2$, and $\omega$ is the fundamental $d$th root of unity.
Therefore, for a rank-one $\sigma_{0}$  the $\bra{n}\!\psi\rangle$'s must fulfill,
\begin{align}\label{rank1 sigma}
&|\!\bra{n}\!\psi\rangle|^2=\bigl(\sigma_{0}\bigr)_{n,n},\nn\\
&\bra{n}\!\psi\rangle\!\bra{\psi}\! n\oplus k\rangle=\bigl(\sigma_{0}\bigr)_{n,n\oplus k}.
\end{align}
The first requirement of Eq.~\eqref{rank1 sigma} implies that the amplitude of the components of the fiducial vector $\psi$ have a certain structure and is determine $(d{-}1)/2$ real parameters $\phi_{d,k}$. The second requirement of Eq.~\eqref{rank1 sigma} could be written as
\beq
\bra{n}\!\psi\rangle\!\bra{\psi}X^k\ket{n}=\bra{n}\frac1{d\sqrt{d+1}}\sum_{j=0}^{d-1}e^{\ii\phi_{j,k}}Z^{jk}\ket{n},
\eeq
or equivalently as
\beq
\ket{\psi}\!\bra{\psi}X^k\bigr|_{\rm diag}=\frac1{d\sqrt{d+1}}\sum_{j=0}^{d-1}e^{\ii\phi_{j,k}}Z^{jk},
\eeq
where the subscript `diag' means  the `diagonal part'. Multiply both sides of the last equation by the (diagonal) operator $Z^{\dagger mk}$ and taking the trace (this operation is allowed since the multiplication of a diagonal operator with any other operator take into account only the diagonal elements of the latter) we obtain
\beq
\bra{\psi}X^kZ^{\dagger mk}\ket{\psi}=\frac1{\sqrt{d+1}}e^{\ii\phi_{m,k}},
\eeq
where we used the fact $\tr{Z^{jk}Z^{\dagger mk}}=d\delta_{j,m}$. The last equation is nothing but the original requirement that $\ket{\psi}$ is a fiducial state for a SIC POM. Here we see that it is actually enough to check $d(d{-}1)/2$ conditions, rather than $d^2{-}1$, as $k{=}1,\ldots,(d{-}1)/2$ and $m{=}0,\ldots,d{-}1$.

\section{Summary and concluding remarks}\label{summary}
To summarize, we have considered the two sets of hermitian operators acting on a $d$-dimensional Hilbert space; the set of SO and the set of MUO. These operators correspond to traceless hermitian operators that form a ($d^2{-}1$)-simplex and $d{+}1$ orthogonal ($d{-}1$)-simplexes in a $d^2-1$ dimension real vector space. For the particular case of prime-power dimension, $d$, we have shown that when the MUO can be considered as points in DAPG and, consequently, the lines on this plane correspond to SO. This relation and its implication was study in particular for the case where either the MUO or SO are rank-one projectors. We have also defined a quasi-probability distribution on the affine planes based on this association. Finally we study the role of HW group elements in prime dimensions as generators of lines in the DAPG when the points are MUO. From this study we were able to obtain a condition for a rank-one SIC POMs.

\begin{acknowledgments}
 We would like to thank Markus Grassl for insightful and stimulating discussions as well a valuable comments on the manuscript. Centre for Quantum Technologies is a Research Centre of Excellence funded by Ministry of Education and National Research Foundation of Singapore. This research was supported in part by NSF Grants No. PHY-1212445.\end{acknowledgments}

\appendix*
\section{The spectrum of MU POM of known SIC POMs in dimensions 2,3,5,7, and 11}
In Secs.~\ref{geometry sub d} and~\ref{hw-geometry sub a} we found the following two properties:
\begin{itemize}
	\item Based on DAPG we can construct (traceless, and trace-one) line operators from point operators. For the case that the former are (rank-one) projectors onto SIC-states, the latter form a complete set ($d{+}1$) MU POM. The spectrum of the POM elements actually determine the rank of the projectors (that is, of the line operators).
	\item In prime dimensions, the HW group elements generate the line operators from a fiducial operator. If the fiducial operator is of rank-one, than the line operators correspond to rank-one SIC POM.
\end{itemize}
In this Appendix we present some results based on a numerical analysis regarding the spectrum of the MU POM that `underpin' known rank-one SIC POMs in prime dimensions 2-11.  

For this aim we should first construct the MU POM from the elements of a (known) SIC POM. Assume that in given (prime) dimension a HW SIC POM generated from a fiducial vector $\ket{\psi_{0}}$
\beq
\ket{\psi_{ab}}\!\bra{\psi_{ab}}= X^{\dagger b}Z^{a}\ket{\psi_{0}}\!\bra{\psi_{0}}Z^{\dagger a}X^{b},
\eeq
exists. Viewing the projectors onto  $\ket{\psi_{a,b}}$ as (trace-one) line operators, we can construct the point operators (the MU POM), by the machinery of DAPG. Following Sec.~\ref{geometry sub d} we denote the $d(d{+}1)$ MU POM by $\tau_{m}^{(j)}$ with $j{=}0,\ldots,d$ and $m{=}0,\ldots,d{-}1$. The probability-operator $\tau_{m}^{(j)}$ corresponds to point $(m,j)$ on the grid of the DAPG. In the DAPG, d lines go through each point, which in terms of the operators is written as Eq.~(\ref{tau in lambda}) with the association of $\mu=(a,b)$ and $\lambda_{a,b}=\ket{\psi_{ab}}\!\bra{\psi_{ab}}$. To give an explicit construction of which lines go through each point, we make the use of the results of Sec.~\ref{hw-geometry sub a}. Let us write the fiducial line operator as a `straight' line, that is as 
\beq
\lambda_0=\sum_{j=0}^d\tau_{0}^{(j)},
\eeq
cf. Fig~\ref{fig:dim3}. Then, according to Sec.~\ref{hw-geometry sub a}, the HW group elements generate all the elements of the SIC POM as lines in the DAPG
\beq\label{lambda ab}
\lambda_{a,b}= \sum_{j=0}^d X^{\dagger b}Z^{a}\tau_{0}^{(j)}Z^{\dagger a}X^{b},
\eeq
and the MU POM are given in terms of the $\lambda$s as
\begin{align}
\tau_{m}^{(d)}&= \frac1{d}\sum_{k=0}^d \lambda_{k,m}\\
\tau_{m}^{(j)}&= \frac1{d}\sum_{k=0}^d \lambda_{m\ominus jk,k},\;j=0,\ldots,d{-}1.
\end{align}
Plugging Eq.~(\ref{lambda ab}) into these equations we get that
\begin{align}
\tau_{m}^{(d)}&= \frac1{d}\sum_{j,k=0}^d X^{\dagger m}Z^{k}\tau_{0}^{(j)}Z^{\dagger k}X^{m}=\frac1{d}X^{\dagger m}\sum_{j,k=0}^d \Bigr(Z^{k}\tau_{0}^{(j)}Z^{\dagger k}\Bigl)X^{m}\\
\tau_{m}^{(j)}&= \frac1{d}\sum_{j,k=0}^d X^{\dagger k}Z^{m\ominus jk}\tau_{0}^{(j)}Z^{\dagger m\ominus jk}X^{k}=\frac1{d}Z^{m}\Bigr(\sum_{j,k=0}^d X^{\dagger k}Z^{\dagger jk}\tau_{0}^{(j)}Z^{jk}X^{k}\Bigl)Z^{\dagger m},\;j=0,\ldots,d{-}1.
\end{align}
Therefore, as was mentioned in Sec.~\ref{geometry sub d}, we see that for a given $j{=}0,\ldots,d$ the probability-operators $\tau_{m}^{(j)}$ with different $m$ values are related by a unitary transformation, hence the spectrum of operators with the same $j$ label is the same. On the other hand, according to this construction probability operators with different $j$ label are not related by a unitary transformation and therefore do not necessarily share the same spectrum.

In matter of fact, we saw on Sec.~\ref{geometry sub d}, that in dimensions 2 and 3 one can there exists a SIC POM for which the spectrum of all $\tau_{m}^{(j)}$ is the same,  $\{\frac1{6}(3\pm\sqrt3)\}$ for the qubit and $\{\frac1{2},\frac1{2},0\}$ for the qutrit. We must note that this spectrum is not unique, that is once can find other spectra that correspond to a SIC POM for a qutrit. For example the spectrum  $\{\frac1{2},\frac1{2},0\}$ for $\tau_{m}^{(d)}$ and $\{\frac1{6},\frac1{6},\frac2{3}\}$ for $\tau_{m}^{(j)}$ $\forall j\neq d$ also correspond to a rank-one SIC POM. The fact that POMs with different $j$ label share the same spectrum may seems to be unexpected, and it is indicates that there is additional symmetry that is not captured by the HW group structure. This feature is not limited to qubits and qutrits SIC POMs. Using conventional numerical routines provided by Wolfarm Mathematica 8, we found that the spectra of the MU POM underpinning the known SIC POMs in prime dimensions 5-11, have the same feature. In what follows, the fiducial vector of the SIC POMs is given in Ref.~\cite{scott10}.\\
\noindent{\bf The spectrum of the probability operators $\tau_{m}^{(j)}$}   
\\\noindent{$d=5$}\\
$$
\begin{array}{ll}
\{0.499925, 0.224729, 0.152916, 0.0930549, 0.0293753\}& \quad \text{for}\; j=0,4,5,\\
\{0.492705, 0.235772, 0.17314, 0.0584088, 0.0399745\}& \quad \text{for}\; j=1,2,3.
\end{array}
$$  
\noindent $d=7$, fiducial state 7a of Ref.~\cite{scott10}\\
$$
\begin{array}{ll}
\{0.285421, 0.285421, 0.285421, 0.0540971, 0.0298802, 0.0298802, 0.0298802\}& \quad \text{for}\; j=7,\\
\{0.419906, 0.150834, 0.150834, 0.150834, 0.0425309, 0.0425309, 0.0425309\}& \quad \text{for}\; j=0,\\
\{0.382799, 0.217579, 0.210489, 0.0908925, 0.0419947, 0.0384167, 0.0178294\}& \quad \text{for}\; j=1,2,4,\\
\{0.425712, 0.177537, 0.116696, 0.0999678, 0.0820381, 0.0814391, 0.0166106\}& \quad \text{for}\; j=3,5,6.
\end{array}
$$  
\noindent $d=7$, fiducial state 7b of Ref.~\cite{scott10}\\
$$
\begin{array}{ll}
\{0.445903, 0.0923495, 0.0923495, 0.0923495, 0.0923495, 0.0923495, 0.0923495\}& \quad \text{for}\; j=7,\\
\{0.284051, 0.284051, 0.284051, 0.0800943, 0.0225843, 0.0225843, 0.0225843\}& \quad \text{for}\; j=0,\\
\{0.410065, 0.172444, 0.157392, 0.137334, 0.0864703, 0.0312058, 0.00508907\}& \quad \text{for}\; j\neq 0,7.
\end{array}
$$  
We see that the SIC POM that is generated form the fiducial state 7b of Ref.~\cite{scott10}, has more symmetry from the one generated by fiducial state 7a of Ref.~\cite{scott10}, in the sense that the more probability-operators have the same spectrum.
\noindent $d=11$, fiducial state 11a of Ref.~\cite{scott10}\\
$$
\begin{array}{l}
\{0.245622, 0.223871, 0.159951, 0.143466, 0.0625246, 0.0568209, 0.0388101, 0.0263934, 0.020796, 0.0154612, 0.00628489\}\\ \text{for}\; j=11,0,3,\\
\{0.226117, 0.218523, 0.208476, 0.104712, 0.0771509, 0.0512401, 0.0488825, 0.047541, 0.0101348, 0.00469396, 0.00252885\}\\ \text{for}\; j=1,5,10,\\
\{0.31832, 0.133805, 0.122566, 0.115196, 0.0861889, 0.0831148, 0.0394999, 0.0371735, 0.0352708, 0.0245194, 0.00434541\}\\ \text{for}\; j=2,4,9,\\
\{0.264926, 0.189422, 0.180079, 0.129948, 0.0699642, 0.0563035, 0.0382599, 0.035081, 0.0164404, 0.0154254, 0.00414998\}\\ \text{for}\; j=6,7,8.
\end{array}
$$  
\noindent $d=11$, fiducial state 11b of Ref.~\cite{scott10}\\
$$
\begin{array}{l}
\{0.298029, 0.180327, 0.14602, 0.0955719, 0.068839, 0.0635472, 0.0597198, 0.0400864, 0.0232912, 0.018484, 0.00608391\}\\ \text{for}\; j=11,0,3,\\
\{0.23682, 0.205874, 0.205481, 0.130529, 0.0623243, 0.0455243, 0.0334152, 0.03105, 0.0227443, 0.0175862, 0.00865247\}\\ \text{for}\; j=1,5,10,\\
\{0.303229, 0.171274, 0.135464, 0.0908191, 0.09051, 0.0627768, 0.056021, 0.0504846, 0.0301582, 0.00716075, 0.00210206\}\\ \text{for}\; j=2,4,9,\\
\{0.323651, 0.134447, 0.121097, 0.0925815, 0.0921314, 0.0826142, 0.0463807, 0.0403358, 0.0264706, 0.0209978, 0.0192941\}\\ \text{for}\; j=6,7,8.
\end{array}
$$  
It is interesting to note that the non-equivalent SIC POMs in dimension 11 that are generated from fiducial states 11a and 11b of Ref.~\cite{scott10}, have the same structure for the spectrum, that is the sets of $j$ labels that have the same spectra is the same for the both SIC POMs. This reflects that these two SIC POMs, though non-equivalent, still share common structure.\vspace{0.3cm}\\
\noindent $d=11$, fiducial state 11c of Ref.~\cite{scott10}\\
$$
\begin{array}{l}
\{0.277642, 0.195466, 0.162084, 0.102926, 0.0788428, 0.0634491, 0.0557729, 0.0222804, 0.0212227, 0.0115049, 0.00880827\}\\ \text{for}\; j=11,0,8,\\
\{0.263093, 0.209056, 0.172451, 0.104608, 0.0841159, 0.0460703, 0.0451983, 0.0332779, 0.021782, 0.0141971, 0.00615008\}\\ \text{for}\; j=1,6,10,\\
\{0.327066, 0.137771, 0.101669, 0.0988332, 0.0870574, 0.0749207, 0.0658295, 0.0339417, 0.0295484, 0.0293712, 0.0139924\}\\ \text{for}\; j=2,7,9,\\
\{0.331579, 0.127224, 0.103875, 0.0903938, 0.0860738, 0.0860713, 0.0483907, 0.0474538, 0.030221, 0.0300363, 0.018681\}\\ \text{for}\; j=3,4,5.
\end{array}
$$ 
The sets of $j$s that share the same spectrum is not the same sets of $j$s for the other two SIC-POMs in dimension 11. However, like its fellow SIC POMs the $j$s are grouped into four groups of three $j$ values.


%
\end{document}